\documentclass[aps,pre,groupaddress,superscriptaddress,reprint,amsmath,amssymb,graphicx]{revtex4-2}
\usepackage{physics}
\usepackage{bm}
\usepackage{bbold}
\usepackage{graphicx}
\usepackage[toc,page]{appendix}
\usepackage{float}
\usepackage{color}
\usepackage{cleveref}
\usepackage{tikz}
\usepackage{xcolor}
\usepackage{booktabs}
\usepackage{amsmath}
\usepackage{mathtools}
\usepackage[normalem]{ulem}
\usepackage{soul}

\begin{document}

\title{Emergence of information interference in stochastic systems\\
with non-diagonal noise and switching environments}
\author{Andrea Marchetti}
\affiliation{Department of Physics and Astronomy ``Galileo Galilei'', University of Padova, Padova, Italy}
\affiliation{INFN, Sezione di Padova, via Marzolo 8, Padova, Italy - 35131}
\author{Daniel Maria Busiello}
\thanks{D.M.B. and G.N. contributed equally to this work, and are listed alphabetically}
\affiliation{Department of Physics and Astronomy ``Galileo Galilei'', University of Padova, Padova, Italy}
\affiliation{Max Planck Institute for the Physics of Complex Systems, Dresden, Germany}
\author{Giorgio Nicoletti}
\thanks{D.M.B. and G.N. contributed equally to this work, and are listed alphabetically}
\affiliation{Quantitative Life Sciences section, The Abdus Salam International Center for Theoretical Physics (ICTP), Trieste, Italy}

\begin{abstract}
\noindent Stochastic forces in natural systems are rarely isotropic. From hydrodynamically coupled colloids to chemical reaction networks, noise contributions are inherently correlated. Together with internal interactions and changing environments, they shape the dependencies between the degrees of freedom of real-world systems, as quantified by their mutual information. In this work, we focus on linearized stochastic systems with both non-diagonal noise matrices and stochastically switching environments. We study how their presence leads to the emergence of information interference, so that the total mutual information cannot be decomposed as the sum of the contributions from deterministic interactions, noise anisotropy, and environmental switching alone. We identify two distinct sources of information interference: a static term, arising from the simultaneous presence of deterministic coupling and noise anisotropy; and a dynamic term, emerging from the interplay between internal processes and environmental switches. We then apply this framework to different physical systems. In the presence of switching temperatures, the mutual information disentangles exactly into internal and environmental contributions. When the noise anisotropy arises instead from hydrodynamic interactions, we find that the presence of a shared fluid can either mask or enhance the information stemming from a non-conservative force depending on its degree of non-reciprocity. Finally, in a fuel-driven chemical reaction network, we show that information interference is controlled by the non-equilibrium driving. These results establish a general information-theoretic perspective on how anisotropic noise and environmental variability shape statistical dependencies in stochastic systems.
\end{abstract}

\maketitle

\section{Introduction}
\noindent From biochemical \cite{Hilfinger2011, thomas2014phenotypic, Tsimring2014, balazsi2011cellular, hyman2014liquid} to ecological \cite{wienand2017fluctuating, wienand2018eco, padmanabha2024spatially, chesson2000mechanisms, hastings2010timescales, poisot2015beyond, kussell2005phenotypic} and neural systems \cite{panzeri2022structures, stein2005neuronal, cecchi2000noise}, there are countless examples of natural systems subjected to noisy, varying and inhomogeneous environments \cite{dorsaz2010diffusion, galanti2014diffusion, chechkin2017brownian}. In most cases, emergent behaviors are equally affected by the specificities of their internal interactions and the properties of the fluctuating external signals \cite{swain2002extrinsic, elowitz2002stochastic, mariani2022disentangling, berton2020thermodynamics, liang2024thermodynamic}. Therefore, any simplistic description that ignores environmental terms falls short in determining the central ingredients governing the responses of systems \cite{chesson1986environmental, rivoire2011value, bowsher2012biochemical, bernhardt2020life, ferrari2018separating, ngampruetikorn2023extrinsic, morrell2024neural}. Similarly, incorporating the effects of the environment into effective couplings might fail to pinpoint those observations that are sheer consequences of the interaction with external variables.

In recent years, information theory has proved to be a useful framework to disentangle, in relatively simple cases, the effects of internal and external couplings \cite{nicoletti2021mutual, nicoletti2022mutual, nicoletti2024interference}. When physical complications appear, such as the presence of temperature gradients, non-linear potentials, or dissipative hidden variables, an exact disentangling might not be possible and interference contributions start emerging. This is quantified by the amount of mutual information that cannot be single-handedly explained by internal or external interactions. Thus, information interference is a genuine consequence of the coupling between the environment and the system's degrees of freedom, and signals when this coupling boosts or masks the dependencies between internal variables \cite{nicoletti2022mutual, moreno2014information}.

However, stochastic systems exhibiting non-diagonal noise terms have been overlooked to date by these recent approaches. Yet, non-diagonal noises naturally emerge in a variety of physical scenarios \cite{kalz2022collisions, fruchart2023odd}. For example, hydrodynamic coupling between two particles immersed in a fluid gives rise to anisotropic diffusion \cite{batchelor1976brownian, Rotne1969, Yamakawa1970, meiners1999direct, das2025unraveling}, torque-dependent self-propulsion in chiral active matter leads to a non-diagonal diffusivity tensor \cite{hargus2021odd, marconi2026emergent}, and internal noise in large diluted chemical systems usually displays cross dependencies in the Fokker--Planck description \cite{van1983stochastic, schmiedl2007stochastic}. In this scenario, information interference can arise from two distinct mechanisms involving environmental variations and non-diagonal contributions. In particular, the latter stems from the fact that anisotropic diffusion induces additional couplings between the relevant degrees of freedom, similarly to what occurs in the presence of a shared environment.

Here, we start by analyzing a linearized stochastic system with non-diagonal noise, introducing all the quantities of interest in a general context. Then, by studying the effect of a switching environment, we show that disentangling is always possible when environmental changes involve only the temperature. Next, to study the role of information interference, we specialize our analysis to the case of hydrodynamic couplings and internal fluctuations in a chemical system. By tinkering with relevant physical parameters, we identify the regimes in which the presence of anisotropy-related terms is beneficial or detrimental to the effective coupling between internal variables. This work opens a new perspective into the study of complex systems, highlighting how an information-theoretic approach might help rationalize observations in terms of timescales and environmental changes, two ingredients that have often received limited attention.

\section{Stochastic systems with non-diagonal noise}
\label{sec:anisotropy}
\noindent We consider a multivariate Ornstein--Uhlenbeck (OU) process for an $N$-dimensional vector of stochastic degrees of freedom (DOFs) $\mathbf{x} = (x_1,\ldots,x_N)^{\mathsf T}$, governed by the Langevin equation
\begin{equation}
\label{eq:Langevin_Ornstein-Uhlenbeck}
    \tau \, \dot{\mathbf{x}}(t) = - \hat{A}\,\mathbf{x}(t)
    + \sqrt{2 \tau}\,\hat{b}\,\boldsymbol{\xi}(t) \;,
\end{equation}
where $\boldsymbol{\xi}(t)$ denotes a vector of independent Gaussian white noises
with zero mean and unit variance, $\hat{A}$ is the interaction matrix, $\hat{b}$ the noise matrix, and $\tau$ the timescale at which the process evolves. In Eq.~\eqref{eq:Langevin_Ornstein-Uhlenbeck}, we incorporate a factor $\tau^{-1/2}$ in the diffusion coefficient to ensure that both drift and diffusion exhibit the same typical timescale. This choice is also compatible with the fluctuation--dissipation theorem in the presence of thermal fluctuations \cite{dabelow2019irreversibility,nicoletti2024tuning,di2024variance,busiello2024unraveling}.
The corresponding Fokker--Planck (FP) equation for the time-dependent probability distribution of $\mathbf{x}$ reads
\begin{equation}
\label{eq:FP_Ornstein-Uhlenbeck}
    \tau \, \dot{p}(\mathbf{x},t) = \nabla\cdot\left[\hat{A}\,\mathbf{x}\,p(\mathbf{x},t) + \hat{B}\,\nabla p(\mathbf{x},t)\right],
\end{equation}
where $\hat{B} = \hat{b}\,\hat{b}^T$ is positive definite if and only if $\hat{b}$ is invertible. We focus on noise covariance matrices $\hat{B}$ that are non-diagonal and therefore induce correlations between different DOFs through their shared stochastic forces. If $-\hat{A}$ is a Hurwitz matrix and $\hat{B}$ is positive definite, Eq.~\eqref{eq:FP_Ornstein-Uhlenbeck} admits a Gaussian steady-state solution $p^\mathrm{st} \sim \mathcal{N}(0, \hat{\Sigma})$, where the covariance matrix $\hat{\Sigma}$ obeys the Lyapunov equation \cite{gardiner}
\begin{equation}
\label{eq:Lyapunov}
    \hat{A}\,\hat{\Sigma} + \hat{\Sigma}\,\hat{A}^T = 2\hat{B} \; .
\end{equation}
Eqs.~\eqref{eq:Langevin_Ornstein-Uhlenbeck}--\eqref{eq:FP_Ornstein-Uhlenbeck} might, for example, represent a system of Brownian particles, where $\mathbf{x}$ denotes particle positions and $\hat{B}$ the diffusion matrix. In this context, a non-diagonal noise is associated with anisotropic diffusion, and we extend the term “anisotropy” to all stochastic systems with non-diagonal noise, even outside diffusion settings.

\subsection{Mutual information and static information interference}
\noindent We now seek to isolate and rationalize the different sources of statistical dependencies that may arise in systems described by Eq.~\eqref{eq:Langevin_Ornstein-Uhlenbeck}. To do so, we consider the stationary mutual information (MI) \cite{ThomasCover2006} between any pair of variables, defined as the Kullback--Leibler divergence
\begin{equation}
\label{eq:MI}
    I_{\mu\nu} = \int dx_\mu dx_\nu \, p_{\mu\nu}^\mathrm{st}(x_\mu, x_\nu)\, \log\left[{\frac{p_{\mu\nu}^\mathrm{st}(x_\mu, x_\nu)}{p_{\mu}^\mathrm{st}(x_\mu)\,p_{\nu}^\mathrm{st}(x_\nu)}}\right]
\end{equation}
where $p_{\mu\nu}^\mathrm{st}$ is the steady-state joint probability distribution of the $\mu$-th and the $\nu$-th DOFs, while $p_{\mu}^\mathrm{st}$ and $p_{\nu}^\mathrm{st}$ are the corresponding marginal distributions (see Appendix~\ref{app:MI_OU} for the explicit formula in the case of Gaussian processes). $I_{\mu\nu}$ measures the statistical dependencies between $x_\mu$ and $x_\nu$ in terms of how much information they share, i.e., of the distance between their joint distribution and the distribution that would arise if the particles were completely independent. Since our results are valid for any pair of particles, in what follows we consider for simplicity $\mathbf{x} = (x_1, x_2)$.

In order to understand how different physical mechanisms contribute to the MI between coupled DOFs, we consider the following approach. Given a system with multiple physical mechanisms that may generate or modulate statistical dependencies, we divide the set of all independently controllable processes into two subsets $\alpha$ and $\beta$. Here, we first focus on \emph{static} processes, i.e., processes that do not induce changes in time of the system's parameters. Then, we compare the full MI to the sum of the contributions that each subset would generate in isolation by considering the \emph{static information interference} between $\alpha$ and $\beta$ as
\begin{equation}
    \label{eq:info_interference_general}
    \Xi_\mathrm{int} = I_{\rm int}
    - \left(I_\alpha + I_\beta \right) \;,
\end{equation}
where $I_{\rm int} = I_{12}$ is the actual MI of the full system, $I_\alpha$ is the MI when only the processes in subset $\alpha$ are present, and analogously for $I_\beta$. Within this framework, the influence of the environment is assumed to be fully mediated by the system parameters, such that static processes correspond to a time-invariant environment. Accordingly, the notation $\Xi_\mathrm{int}$ emphasizes that any observed static information interference arises solely from \emph{internal} system mechanisms, i.e., mechanisms that are independent of environmental dynamics and do not involve temporal variations in external conditions. The notation $I_{\rm int}$ highlights the same for the MI of the whole system. A positive (negative) $\Xi_\mathrm{int}$ indicates that the two mechanisms $\alpha$ and $\beta$ combine \emph{constructively} (\emph{destructively}): the actual MI is larger (smaller) than what one would expect by simply adding their individual contributions. In this work, we apply this framework to different subsets of processes, identifying in each case the relevant $\alpha$ and $\beta$.

\subsection{Anisotropy-driven information interference}
\noindent As a first application of the framework introduced above, we consider systems in which both the deterministic coupling between particles and the noise anisotropy are independently tunable. For ease of interpretation, we decompose the drift and diffusion matrices into their diagonal and non-diagonal components,
\begin{equation}
    \hat{A} = \hat{A}_\mathrm{d} + \hat{A}_\mathrm{nd}, \quad
    \hat{B} = \hat{B}_\mathrm{d} + \hat{B}_\mathrm{nd}
\end{equation}
where $\hat{A}_\mathrm{d}$ and $\hat{B}_\mathrm{d}$ are diagonal matrices representing the self-relaxation and 
self-diffusion coefficients, respectively. $\hat{A}_\mathrm{nd}$ quantifies instead the direct deterministic coupling between the particles, and $\hat{B}_\mathrm{nd}$ the noise correlations due to anisotropy, with $(\hat A_\mathrm{nd})_{ii} = (\hat B_\mathrm{nd})_{ii} = 0$. When the off-diagonal parts of both matrices vanish, the two DOFs are fully decoupled and $I_{\rm int} = 0$ identically.

Since the two mechanisms are taken to be independent in this example, each can be switched off separately, giving rise to two reference MIs: $I_\mathrm{coup} = I_{\rm int}\big|_{\hat{B}_\mathrm{nd}=0}$, the MI due solely to the deterministic coupling; and $I_\mathrm{an} = I_{\rm int}\big|_{\hat{A}_\mathrm{nd}=0}$, the MI generated purely by the noise anisotropy. Both correspond to physically realizable limits. By identifying $\alpha = \mathrm{coup}$ and $\beta = \mathrm{an}$ in Eq.~\eqref{eq:info_interference_general}, the information interference reads
\begin{equation}
    \label{eq:MI_decomposition_anisotropy}
    \Xi_\mathrm{int}(\hat A, \hat B) = I_{\rm int}(\hat A, \hat B)  - \left( I_\mathrm{coup}(\hat A, \hat B_\mathrm{d}) + I_\mathrm{an}(\hat A_\mathrm{d}, \hat B)\right).
\end{equation}
Thus, $\Xi_\mathrm{int}$ quantifies the non-additive interplay stemming from the simultaneous presence of deterministic couplings and noise anisotropy.

For illustrative purposes, we consider the following simple example:
\begin{equation}
\label{eq:A_b}
    \hat{A} = \begin{pmatrix}
        1 & -g \\
        -g & 1
    \end{pmatrix}, ~~~~~\hfill \hat{b} = \begin{pmatrix}
        \sqrt{D} & \sqrt{a} \\
        \sqrt{a} & \sqrt{D}
    \end{pmatrix}
\end{equation}
with $g$ the internal coupling between the particles, $D > 0$ the self-diffusion coefficient, and $a$ quantifying the anisotropy of the noise. The Hurwitz condition on $-\hat{A}$ corresponds to requiring that $|g|<1$, and for $\hat{b}$ to be invertible, we require $D \neq a$. The corresponding noise covariance matrix reads
\begin{equation}
    \hat{B} = \begin{pmatrix}
        D + a & 2 \sqrt{a D} \\
        2 \sqrt{a D} & D + a
    \end{pmatrix}
\end{equation}
so that any $a>0$ will result in anisotropic diffusion.

The MI due solely to the deterministic coupling is obtained by setting $a = 0$,
\begin{equation}
\label{eq:MI_drift}
    I_\text{coup}(\hat A) = \frac{1}{2}\log\left(\frac{1}{1 - g^2}\right),
\end{equation}
while that generated purely by the noise anisotropy is given by setting $g = 0$,
\begin{equation}
\label{eq:MI_noise}
    I_\text{an}(\hat B) = \log\left(\frac{D + a}{D - a}\right). 
\end{equation}
Eqs.~\eqref{eq:MI_drift}--\eqref{eq:MI_noise} can be directly computed by solving Eq.~\eqref{eq:Lyapunov} and using the expression outlined in Appendix \ref{app:MI_OU}. We stress that, whenever non-diagonal terms appear in $\hat{b}$, the MI is non-zero even if no deterministic couplings are present, due to the noise-induced correlations. In the example presented, $I_\mathrm{coup}$ is independent of $\hat B_\mathrm{d}$ and $I_\mathrm{an}$ of $\hat A_\mathrm{d}$ purely because the diagonal elements of each matrix are identical. When both processes are present, their contributions combine in a non-trivial way and generate an information interference quantified by
\begin{equation}
\label{eq:MI_interference}
\Xi_\mathrm{int}(\hat A, \hat B) = \log\left(1 + \frac{2g\sqrt{aD}}{D + a}\right) \;.
\end{equation}
The appearance of the mixed term $g\sqrt{aD}$ in Eq.~\eqref{eq:MI_interference} signals that the dependencies stemming from the deterministic coupling and the noise anisotropy cannot be disentangled -- correlated noise fundamentally modifies how deterministic interactions contribute to the statistical dependencies between the particles. In particular, we can distinguish two cases. When $g > 0$, the internal interference is constructive ($\Xi_\mathrm{int} > 0$), so that the total MI is larger than the sum of the individual contributions. When $g<0$, interference is destructive ($\Xi_\mathrm{int} < 0$), and the combined effect of drift and noise decreases the overall dependencies between the two DOFs.

\subsection{Switching environment and dynamic information interference}
\label{sec:switching_environment}
\noindent Having characterized the information interference between internal coupling mechanisms, we now introduce an external source of statistical dependencies: a stochastically switching environment \cite{nicoletti2021mutual, nicoletti2022mutual}. In particular, we assume that the DOFs share an environment that switches stochastically between a set of $S$ discrete states, and that each state may in general define different drift and noise matrices. We rewrite Eq.~\eqref{eq:Langevin_Ornstein-Uhlenbeck} as
\begin{equation}
\label{eq:Langevin_switch}
    \tau\,\dot{\mathbf{x}}(t) = -\hat{A}_{i(t)}\,\mathbf{x} + \sqrt{2\tau}\,\hat{b}_{i(t)}\,\bm{\xi}(t),
\end{equation}
with $i(t) \in \{1, \dots, S\}$ denoting the environmental state at time $t$. Defining $\tau_\text{jumps}$ as the timescale of the environmental process, the combined system -- i.e., internal DOFs plus the environment -- is described by the FP equation
\begin{equation}
\label{eq:ME_switch}
\begin{aligned}
    \dot{p}_i(\mathbf{x},t) =& \frac{1}{\tau}\nabla\cdot\left[\hat{A}_i\,\mathbf{x}(t)\,p_i(\mathbf{x},t) + \hat{B}_i\,\nabla p_i(\mathbf{x},t)\right] + \\
    +&\frac{1}{\tau_{\text{jumps}}}\sum_j \Big[W_{ij}\,p_j(\mathbf{x},t) - W_{ji}\,p_i(\mathbf{x},t)\Big],
\end{aligned}
\end{equation}
where $p_i(\mathbf{x},t)$ is the probability density of the DOFs in the $i$-th environmental state, and $W_{ij}/\tau_\text{jumps}$ is the switching rate from state $j$ to state $i$.

At variance with the previous case, here we are considering \emph{dynamic} processes, i.e., processes in which the environment induces time dependence in the system parameters. For each environmental state, and in analogy with the notation introduced above, we denote by $I_{\mathrm{int},i} = I_{\mathrm{int}}(\hat{A}_i, \hat{B}_i)$ the MI arising solely from internal mechanisms -- namely, the value obtained in the corresponding static environment associated with the $i$-th environmental state.

On top of internal mechanisms, the presence of a switching environment affecting all DOFs introduces additional statistical dependencies. To account for these dependencies, we introduce the MI induced purely by the shared switching environment, $I_\mathrm{env}$. This is obtained by setting to zero the deterministic coupling between the particles and the noise anisotropy -- i.e., all the internal interaction mechanisms. On the other hand, in the absence of switching, we are left with $S$ internal MI terms, $I_{\mathrm{int},i}$. Following previous work \cite{nicoletti2022mutual}, in the case in which the environmental switching process is much slower than the internal dynamics ($\tau_{\text{jumps}} \gg \tau$), we introduce the \emph{dynamic} information interference between internal and environmental processes as
\begin{equation}
    \label{eq:MI_decomposition_env}
    \Xi_\mathrm{int,env}
    = I_{\rm tot} - \left(
    \langle I_\mathrm{int} \rangle + I_\mathrm{env} \right),
\end{equation}
where
\begin{equation}
\label{eq:Iint}
    \langle I_\text{int}\rangle = \sum_i \pi_i^{\rm st}\, I_{\text{int},i}
\end{equation}
is the internal MI averaged over the stationary probability distribution of the $S$ environmental states, while $I_{\rm tot} = I_{12}$ denotes the total MI computed for the full system according to Eq. \eqref{eq:MI}. Notably, $\langle I_\text{int}\rangle$ does not carry information on the dynamics of the environment, but only on its stationary statistics. Whenever $\Xi_\mathrm{int,env} = 0$, the internal and environmental contributions are fully disentangled, that is, the statistical dependencies induced by a changing environment can -- in principle -- be fully separated from those due to internal interactions \cite{nicoletti2021mutual, nicoletti2022mutual, nicoletti2024interference}.

\subsubsection{Slow-jump limit}
\label{sec:slow_jumps}
\noindent Here, we give mathematical ground for the decomposition introduced in Eq.~\eqref{eq:MI_decomposition_env} for the limit case of a slowly switching environment ($\tau_{\text{jumps}} \gg \tau$).

In this slow-jump limit, the timescale separation solution gives $p_i(\mathbf{x},t) \simeq \pi_i(t)\, P^\mathrm{st}_i(\mathbf{x})$ \cite{nicoletti2024gaussian, nicoletti2024information, nicoletti2025stochastic}, where $\pi_i(t)$ obeys the master equation
\begin{equation}
    \dot{\pi}_i(t) = \tau_\text{jumps}\,\sum_j \Big[{W}_{ij}\,\pi_j(t) - {W}_{ji}\,\pi_i(t)\Big]
\end{equation}
while $P^\mathrm{st}_i \sim \mathcal{N}(0, \hat{\Sigma}_i)$ is the solution of the steady-state FP equation
\begin{equation}
\label{eq:FP_internal_slow_switching}
    0 = \nabla\cdot\left[\hat{A}_i\,\mathbf{x}(t)\,P_i(\mathbf{x}) + \hat{B}_i\,\nabla P_i(\mathbf{x})\right] \; .
\end{equation}
Thus, $\hat{\Sigma}_i$ is the covariance matrix solving the Lyapunov equation (Eq.~\eqref{eq:Lyapunov}) with the drift and noise matrices associated with the $i$-th environmental state, 
\begin{equation}
\label{eq:Lyapunov_mixture}
    \hat{A}_i\hat{\Sigma}_i + \hat{\Sigma}_i(\hat{A}_i)^T = 2\hat{B}_i \, ,
\end{equation}
whereas the joint distribution of internal DOFs, $P^\mathrm{st}(\mathbf{x}) = \sum_i \pi_i^\mathrm{st}\, P^\mathrm{st}_i(\mathbf{x})$, is a Gaussian mixture distribution. 

In general, the MI of a Gaussian mixture cannot be computed in closed form. Therefore, we resort to finding optimal bounds on the entropy of Gaussian mixtures \cite{kolchinsky2017estimating}, which can then be used to bound the corresponding MI \cite{nicoletti2021mutual}. If $P^{\rm st}_{12,i}$ is the solution of Eq.~\eqref{eq:FP_internal_slow_switching} and $P^{\rm st}_{1,i}$, $P^{\rm st}_{2,i}$ are the associated marginal probability distributions, these bounds read 
\begin{equation}
\label{eq:MI_bounds}
    I_{\rm tot}^\text{slow,up/low} = \langle I_\text{int}\rangle - \sum_i \pi_i \log \frac{\prod_{\mu = 1}^2\sum_j \pi_j e^{-d_1(P^{\rm st}_{\mu,i} || P^{\rm st}_{\mu,j})}}{\sum_j \pi_j e^{-d_2(P^{\rm st}_{12,i} || P^{\rm st}_{12,j})} }\;,
\end{equation}
where $d_1$ is the Chernoff-$\alpha$ divergence and $d_2$ the Kullback-Leibler divergence \cite{ThomasCover2006} for the lower bound, and the opposite for the upper bound. Thus, the bounds suggest that $\langle I_\text{int}\rangle$ is the most suitable choice to combine the MI $I_{\mathrm{int},i}$ associated with each $P^{\rm st}_{12,i}$.
 
Notably, it is easy to check that, when the switching involves only a scalar prefactor that multiplies both the drift and diffusion matrices -- i.e., $\hat{A}_i = k_i \hat{A}$ and $\hat{B}_i = k_i \hat{B}$ -- the total MI coincides with the internal one, since the Lyapunov equation does not depend on the environmental state (see Appendix~\ref{app:fast_switching_prefactors}). 

\subsubsection{Fast-jump limit}
\noindent Here, we briefly comment on the other limit case of a fast switching environment ($\tau_\text{jumps} \ll \tau$), where the stochastic variables effectively experience the environment as averaged over its stationary distribution. The timescale separation limit gives $p_i(\mathbf{x},t) \simeq \pi^\mathrm{st}_i P(\mathbf{x},t)$ \cite{nicoletti2021mutual, nicoletti2024information}, where $\pi^\mathrm{st}_i$ is the stationary probability of the jump process, i.e., it satisfies
\begin{equation}
    0 = \tau_\text{jumps}\,\sum_j \Big[{W}_{ij}\,\pi^{\rm st}_j(t) - {W}_{ji}\,\pi^{\rm st}_i(t)\Big]
\end{equation}
for all $i$ and $j$, while $P(\mathbf{x},t)$ obeys
\begin{equation}
\label{eq:FP_fast_switching}
    \dot{P}(\mathbf{x},t) = \nabla\cdot\left[\langle\hat{A}\rangle\,\mathbf{x}(t)\,P(\mathbf{x},t) + \langle\hat{B}\rangle\,\nabla P(\mathbf{x},t)\right],
\end{equation}
with $\langle\cdot\rangle$ denoting the average over the statistics of the environmental states. In other words, in the fast-jump limit, the solution is the same as that of a Langevin system with effective drift and diffusion matrices given by the mean drift and diffusion matrices over all possible environmental states. Thus, the total MI is $I_{\rm tot}^\mathrm{fast} = I_\mathrm{int}(\langle\hat{A}\rangle, \langle\hat{B}\rangle)$, showing that the dynamics of the environment does not generate any additional statistical dependencies between the particles. Consequently, we will focus on the slow-jump limit in the rest of this work.

\begin{figure}
    \centering
    \includegraphics[width=\columnwidth]{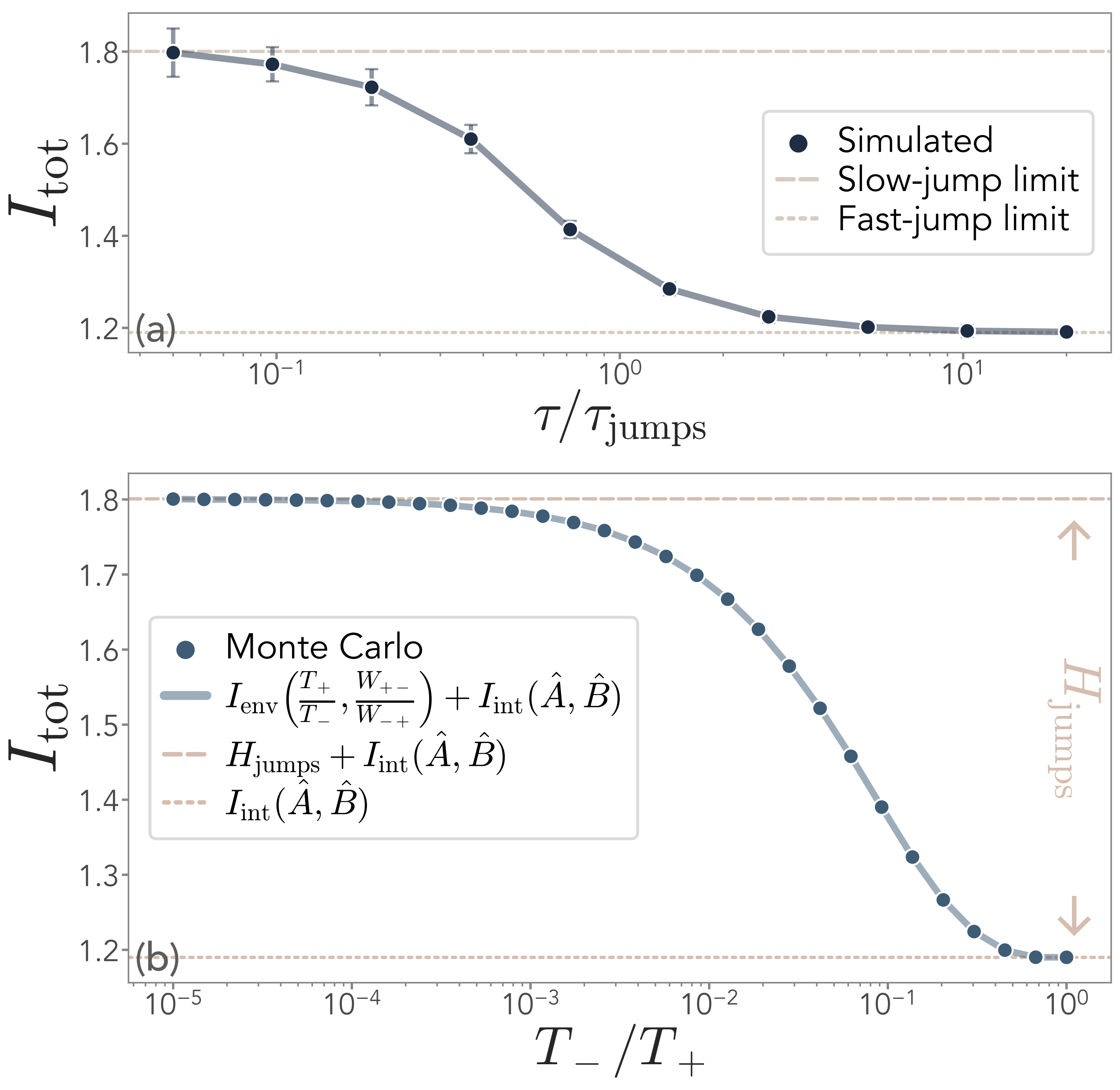}
    \caption{(a) Total MI between two DOFs $x_1$ and $x_2$ estimated with a k-nearest neighbour estimator \cite{kraskov2004estimating} from simulated trajectories of the Langevin equations (Eq.~\eqref{eq:Langevin_switch}), as a function of the ratio between the timescale $\tau$ of the DOFs and the timescale $\tau_\mathrm{jumps}$ of the environmental jumps between a small temperature ($T_- = 10^{-5}$) and a high temperature ($T_+ = 1$). When $\tau \gg \tau_\mathrm{jumps}$, the environment is much faster than the DOFs and the MI converges to a value that only depends on the noise and drift matrices in the various environmental states. As the environment becomes slower, the MI increases due to the additional dependencies induced by the shared temperature switch and converges to the value predicted by the slow-jump timescale separation limit. (b) In the slow-jump limit, $I_{\rm tot}$ depends on the ratio between the two temperatures, $T_-/T_+$. In particular, $I_{\rm tot}$ can be disentangled into an internal term that only depends on the noise and diffusion matrices, $I_\mathrm{int}(\hat{A},\hat{B})$, and an environmental term that only depends on the temperature jumps, $I_\mathrm{env}(T_-/T_+,W_{+-} / W_{-+} )$. The latter converges to the jump entropy $H_\mathrm{jumps}$ as $T_-/T_+ \to 0$. In this figure, $D = 1$, $a = 0.1$, $g = 1/2$, $W_{+-} = 0.7$, $W_{-+} = 0.3$.}
    \label{fig:T_switch}
\end{figure}

\section{Disentangling interactions from switching environmental temperatures}
\label{sec:switching_temperatures}
\noindent While the computation of the divergences appearing in Eq.~\eqref{eq:MI_bounds} is often cumbersome, simple expressions can be obtained in specific cases and appropriate limits. Here, we extend the derivation made in \cite{nicoletti2021mutual} for a system with switching temperatures to the case of a general noise matrix. Specifically, we consider a system that jumps between two temperatures $T_+$ and $T_-$ with rates $W_{+-}$ and $W_{-+}$. Accordingly, the noise matrix switches between two states $\hat{B}_\pm = T_\pm\,\hat{B}$, with $\hat{B}$ being a diffusion matrix that does not depend on the environment. In Figure \ref{fig:T_switch}a, we show the MI obtained numerically from the Langevin equations for different values of $\tau/\tau_\mathrm{jumps}$.

When $\tau/\tau_\mathrm{jumps}$ is large, the environment is fast and the MI converges to the one stemming only from the drift matrix and the temperature-independent diffusion matrix, $I_\text{int}(\hat{A}, \hat{B})$ -- this is due to the fact that the MI does not depend on the temperature, whence $I_{\rm int}(\hat A, \langle T\rangle \hat B) = I_{\rm int}(\hat A, \hat B)$. However, as the environment becomes slower, the MI increases due to the additional dependencies induced by the shared temperature switch. When $\tau/\tau_\mathrm{jumps}$ is small, the MI, according to Eq.~\eqref{eq:MI_bounds}, can be bounded as
\begin{equation}
\label{eq:MI_bounds_temperature}
    I_{\rm tot}^\text{slow,up/low} = I_\text{int}(\hat{A}, \hat{B}) + I_\text{env}^\text{up/low}\left(\frac{T_+}{T_-}, \frac{W_{+-}}{W_{-+}}\right).
\end{equation}
Crucially, the bounds greatly simplify in the limits in which the two temperatures are very close ($T_+ \approx T_-$) or very different ($T_+ \gg T_-$). Moreover, in these two regimes, upper and lower bounds converge to the same value, enabling the exact quantification of the MI and the identification of a vanishing information interference ($\Xi_\mathrm{int,env} = 0$ in Eq.~\eqref{eq:MI_decomposition_env}):
\begin{equation}
\label{eq:MI_slow_temperature_limits}
    I_{\rm tot}^\text{slow} = \begin{cases}
        I_\text{int}(\hat{A}, \hat{B}) + H_\text{jumps} & \text{if}~~ T_+ \gg T_- \\
        I_\text{int}(\hat{A}, \hat{B}) \hfill & \text{if}~~ T_+ \approx T_-,
    \end{cases}
\end{equation}
with $H_\text{jumps} = -\sum_i\pi_i^{\rm st}\,\log\pi_i^{\rm st}$ being the entropy of the environmental jump process and playing the role of $I_{\rm env}$. For intermediate values of the temperature ratio, the MI must be computed numerically by means of Monte Carlo simulations \cite{nicoletti2021mutual}. In Figure \ref{fig:T_switch}b, we show that the MI always disentangles in form as
\begin{equation}
\label{eq:T_disentangling}
    I_{\rm tot}^\text{slow} = I_\text{int}(\hat{A}, \hat{B}) + I_\mathrm{env}\left(\frac{T_+}{T_-}, \frac{W_{+-}}{W_{-+}}\right)
\end{equation}
where $I_\mathrm{env}$ is, as before, the MI due to the temperature jumps in the absence of internal couplings. As a final remark, we note that a switching prefactor affecting only $\hat A$ is, from the perspective of the MI, equivalent to a switching temperature, once such a prefactor is identified as an inverse temperature. We refer to Appendix~\ref{app:slow_switching_temperature} for a detailed derivation of these results.

\section{Brownian particles with hydrodynamic couplings}
\label{sec:hydro}
\noindent We now apply the ideas introduced in Section \ref{sec:anisotropy} to a system of two Brownian particles subject to an external force and immersed in a shared fluid. The presence of a common solvent gives rise to hydrodynamic interactions, which are long-range interactions originating from the disturbance of the fluid flow induced by each particle \cite{Rotne1969}. The emergence of hydrodynamic couplings between two colloidal particles in a solution has been studied both theoretically and experimentally \cite{meiners1999direct, Berut2014, Berut2016}, and recently through information theory \cite{das2025unraveling}. 
Furthermore, whenever these particles are trapped by optical tweezers, the non-conservative components of the external forcing lead to non-equilibrium dynamics \cite{Wu2009}.

We consider two spherical Brownian particles of radii $r_i$, trapped by a confining force so that their equilibrium positions are separated by a distance $d$. They are immersed in a fluid in thermal equilibrium at temperature $T$, which provides both viscous dissipation and thermal fluctuations satisfying the fluctuation--dissipation theorem. For simplicity and clarity of exposition, we focus on particles moving in a one-dimensional system. In the low Reynolds number regime, where inertial effects are negligible, the particle motion is accurately described by an overdamped Langevin equation with hydrodynamic interactions, encoded in a symmetric, positive-definite mobility matrix $\hat{\mu}$ \cite{Ermak1978}. If $\mathbf{x} = (x_1,x_2)$ represents the displacement of the two particles from their equilibrium positions, in a static environment the system obeys
\begin{equation}
\label{eq:Langevin_hydro}
    \tau \, \dot{\mathbf{x}}(t) = \hat{\mu}\,\mathbf{F}(\mathbf{x}(t))
    + \sqrt{2 \tau\,k_B\,T}\,\hat{\sigma}\,\boldsymbol{\xi}(t),
\end{equation}
where $\mathbf{F}(\mathbf{x})$ denotes the external force acting on the particles, $k_B$ the Boltzmann's constant, $\hat{\mu}$ the mobility matrix, and $\hat{\sigma}\,\hat{\sigma}^T = \hat{\mu}$. In the presence of hydrodynamic interactions, $\hat{\mu}$ can be approximated by the Rotne--Prager--Yamakawa (RPY) tensor \cite{Rotne1969, Yamakawa1970}. For two particles with different radii that move in one dimension and that cannot overlap, the RPY tensor reads \cite{Zuk2014}
\begin{equation}
\label{eq:RPY}
        \hat{\mu} =  \frac{1}{6\pi\eta} \begin{pmatrix}
        1/r_1 & \epsilon \\
        \epsilon & 1/r_2
    \end{pmatrix}
\end{equation}
with $\eta$ the viscosity of the fluid, and
\begin{equation}
\label{eq:RPY_epsilon}
    \epsilon =  \dfrac{3}{2\tilde d}\left(1 - \dfrac{r_1^2 + r_2^2}{3\tilde d^2}\right)
\end{equation}
with $\tilde d > r_1 + r_2$ denoting the interparticle distance. In principle, as the particles move, their distance changes over time, and the mobility matrix is a function of the particles' positions. However, in Eqs. \eqref{eq:Langevin_hydro}--\eqref{eq:RPY_epsilon} we assume that the particles' displacements are small compared to the distance between their equilibrium positions ($x_i \ll d$) \cite{Berut2014, Berut2016}, so that we can effectively consider the particle distance to be constant and confuse $\tilde d \approx d$. By reducing the mobility matrix to a constant, this greatly simplifies the problem and allows for analytical treatment. 

In this setting, the two independent static processes at play are the action of the external force $\mathbf{F}$ and hydrodynamic interactions -- represented by $\hat{\mu}$. Crucially, the latter affects both the anisotropy of the noise and the overall drift matrix.

\subsection{Conservative forces hinder fluid properties}
\label{sec:hydro_conservative}
\noindent We first consider the case of a conservative external  force, so that the FP equation associated with Eq.~\eqref{eq:Langevin_hydro} can be written as
\begin{equation}
\label{eq:FP_conservative}
    \tau \, \dot{p}(\mathbf{x},t) = \nabla\cdot\Big[\hat{\mu}\,\nabla U(\mathbf{x})\,p(\mathbf{x},t) + k_B T\,\hat{\mu}\,\nabla p(\mathbf{x},t)\Big]
\end{equation}
with $\mathbf{F}(\mathbf{x}) = -\nabla U(\mathbf{x})$ for some potential $U(\mathbf{x})$.
We immediately have that, at stationarity and with appropriate boundary conditions, the dependence on the fluid -- i.e., on $\hat{\mu}$ -- trivially cancels out. Indeed, the resulting steady-state probability distribution is the equilibrium solution
\begin{equation}
\label{eq:hydro_conservative}
    {p}^\mathrm{st}(\mathbf{x})  = \frac{e^{-U(\mathbf{x})/k_B T}}{\int dx_1 dx_2\, e^{-U(\mathbf{x})/k_B T}} \;.
\end{equation}
Since this expression does not depend on $\hat{\mu}$, the associated MI will be a function of $U(\mathbf{x})$ and the temperature $T$ only (reducing to a function only of $U(\mathbf{x})$ in the case of a harmonic potential). This reveals that, in hydrodynamically interacting systems in a potential, the additional fluid-mediated coupling does not give rise to any statistical dependencies, since they solely arise from the external force acting on the particles.

\subsection{Fluid-mediated interactions may be detrimental}
\begin{figure}
    \centering
    \includegraphics[width=\columnwidth]{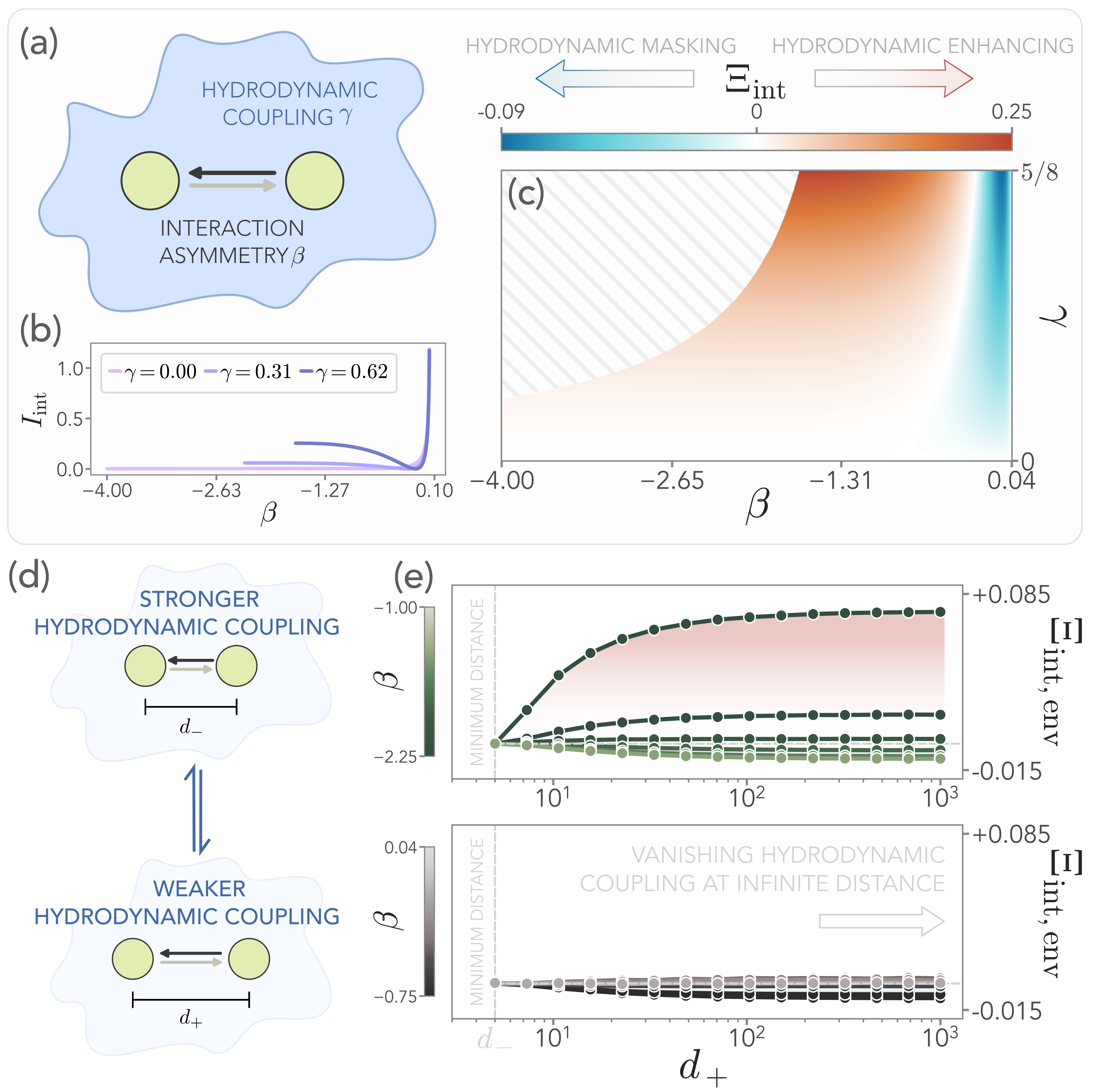}
    \caption{(a) Two identical particles ($r_1 =r_2 = r = 1$) immersed in a shared fluid, which induces a hydrodynamic coupling regulated by the dimensionless parameter $\gamma$. The particles interact non-reciprocally through the external force, with $\beta$ quantifying the interaction asymmetry. (b) As $\gamma$ increases, so does the MI between the particles $I_\mathrm{int}$, with a non-monotonic behavior as a function of $\beta$. (c) The presence of the fluid induces a non-diagonal noise matrix and leads to the emergence of static information interference $\Xi_\mathrm{int}$. Such interference can be both constructive, signalling that the hydrodynamic coupling is enhancing information, or destructive, when it masks information. We require that $\gamma \in (0,5/8)$, which is equivalent to requiring that $d > 2r$ in the RPY tensor (Eq.~\eqref{eq:RPY}). (d-e) When the distance between the particles switches between two values $d_+$ and $d_-$, a dynamic information interference term $\Xi_\mathrm{int, env}$ arises so that the total MI is not the sum of the internal one -- due to the couplings -- and the environmental one -- due to the switches. The interference term is positive for negative enough values of $\beta$, while it becomes negative and eventually vanishes as $\beta$ increases. Unless otherwise specified, in this Figure $\alpha_1 = \alpha_2 = 1/5$, $T = 1$, $\eta = 1$, $d_- = 5r$. Notice that, for these parameters, the system is stable for positive $\beta$ such that $\beta < 1/25$. The region of linear instability for negative $\beta$ is shown for each figure panel.}
    \label{fig:hydro}
\end{figure}

\noindent We now move to a case where the force is linear but non-conservative, that is, $\mathbf{F}(\mathbf{x}) = -\hat{K}\mathbf{x}(t)$ with non-symmetric $\hat{K}$. In this case, the FP equation associated with Eq.~\eqref{eq:Langevin_hydro} is akin to Eq.~\eqref{eq:FP_Ornstein-Uhlenbeck}, and the steady-state solution is Gaussian. The total MI due to internal processes only depends on the dimensionless parameters $\gamma = r_1\epsilon$, $R_r = r_1/r_2$, $\alpha_i = K_{ii}/K_{12}$, and $\beta = K_{21}/K_{12}$, and is given by
\begin{align}
\label{eq:MI_internal_hydro}
    I_{\rm int}(\hat{\mu}, \hat{K}) & = \frac{1}{2}
\log\left[\frac{\alpha_2(f_1+\alpha_1 R_r)-R_r f_3}{\alpha_1\alpha_2-\beta}\right] \\
& + \frac{1}{2}
\log\left[\frac{\beta f_3+\alpha_1(f_2+\alpha_1 R_r)}{(f_1+\alpha_1 R_r)(f_2+\alpha_1 R_r)+R_r f_3^2}\right] \nonumber
\end{align} 
where $f_1 = \alpha_2 + 2\gamma$, $f_2 = \alpha_2 + 2\beta\gamma$, and $f_3 = \beta - 1$. The parameter $\beta$ quantifies the asymmetry or non-reciprocity of the interactions (see Figure~\ref{fig:hydro}a), with $\beta = 1$ recovering the conservative case. Notice that the MI diverges as the system approaches the edge of linear stability for $\beta > 0$ \cite{barzon2025excitation} (see Figure~\ref{fig:hydro}b).

Similarly to what done in the previous section, we decompose the force coupling matrix and the mobility matrix into their diagonal and non-diagonal parts, $\hat{K} = \hat{K}_\mathrm{d} + \hat{K}_\mathrm{nd}$ and $\hat{\mu} = \hat{\mu}_\mathrm{d} + \hat{\mu}_\mathrm{nd}$. Then, the MI due only to the force-induced couplings, $I_\mathrm{force}$, is obtained from Eq.~\eqref{eq:MI_internal_hydro} by setting $\epsilon = 0$, so that $\hat{\mu}_\mathrm{nd}$ vanishes:
\begin{equation}
\label{eq:MI_hydro_nonconservative_drift}
    I_\mathrm{force}(\hat{K}, \hat{\mu}_\mathrm{d}) = \frac{1}{2}
\log\left[\frac{\alpha_2f_4- R_rf_3}{\alpha_1\alpha_2-\beta}\,\frac{\beta f_3+\alpha_1f_4}{f_4^2+R_rf_3^2}\right]
\end{equation}
with $f_4 = \alpha_2+\alpha_1 R_r$. We notice that, if the particles are identical, $R_r = 1$ and $I_\mathrm{force}$ does not depend on $\hat\mu_\mathrm{d}$ any longer. Conversely, the MI due only to fluid-induced couplings, $I_\mathrm{fluid}$, is obtained when the elements of $\hat{K}_\mathrm{nd}$ are zero, so that the force coupling matrix $\hat{K}$ is diagonal. However, this case is trivially equivalent to that of a conservative force described in Section \ref{sec:hydro_conservative} with $U(\mathbf{x}) = \frac{1}{2}\mathbf{x}^T\hat{K}_\mathrm{d}\mathbf{x}$, so that $I_\mathrm{fluid}$ does not depend on $\hat{\mu}$. Since  $\hat{K}_\mathrm{d}$ is diagonal, we immediately have that the joint stationary distribution in Eq.~\eqref{eq:hydro_conservative} is factorized, implying that $I_\mathrm{fluid} = 0$.

Overall, the static information interference between the two internal processes can be written as
\begin{equation}
    \Xi_\mathrm{int}(\hat{K},\hat{\mu}) = I_{\rm tot}(\hat{K},\hat{\mu}) - I_\mathrm{force}(\hat{K}, \hat{\mu}_\mathrm{d})
\end{equation}
according to the notation of Eq.~\eqref{eq:info_interference_general}. As a consequence, $\Xi_\mathrm{int}$ effectively quantifies how the presence of the fluid alters the dependencies between the particles induced by the external force. In Figure \ref{fig:hydro}c, we focus on the case of identical particles and show that $\Xi_\mathrm{int}$ can be both greater or smaller than zero depending on the value of the parameters $\beta$ and $\gamma$. In particular, when $\beta$ is large and negative -- i.e., when the force-induced couplings are highly non-reciprocal and of opposite sign -- hydrodynamic interactions enhance the information between the particles beyond what is provided by the force-induced couplings alone. Conversely, small values of $\beta$ lead to hydrodynamic masking, with the presence of the fluid diminishing the overall dependence.

\subsection{Information interference from switching mobility with nonconservative forces}
\noindent We now consider the case of two particles whose equilibrium distance switches between two values $d_+$ and $d_-$ (Figure \ref{fig:hydro}d). In this case, the mobility matrix changes between the values
\begin{equation}
\label{eq:mobility_stiching}
        \hat{\mu}_\pm =  \frac{1}{6\pi\eta} \begin{pmatrix}
        1/r_1 & \epsilon_\pm \\
        \epsilon_\pm & 1/r_2
    \end{pmatrix}
\end{equation}
where the values of $\epsilon_\pm$ can be deduced from Eq.~\eqref{eq:RPY_epsilon}.

We focus on the limit of slow jumps between the particles' equilibrium distances, since in the fast-jump limit the MI depends only on the mobility matrix averaged over the environmental states. In this limit, the interplay between environmental switching and internal processes gives rise to dynamic information interference, $\Xi_\mathrm{int, env}$. As before, the steady-state solution is a Gaussian mixture $P^\mathrm{st}(\mathbf{x}) = \sum_i \pi_i^\mathrm{st}\, P^\mathrm{st}_i(\mathbf{x})$, whose components $P^\mathrm{st}_i$ solve Eq.~\eqref{eq:FP_internal_slow_switching}. Clearly, in the absence of internal couplings, and therefore of hydrodynamic interactions, environmental switching cannot induce any form of statistical dependence, as the dynamics of the particles no longer depends on their distance. As such, $I_\mathrm{env} = 0$, and the dynamic information interference in Eq.~\eqref{eq:MI_decomposition_env} simplifies to the difference between the total MI and the average of the internal MI over the various environmental states: 
\begin{equation}
\label{eq:xi_hydro}
    \Xi_\mathrm{int, env} (\hat{\mu}_\pm, \hat{K}) = I_{\rm tot}(\hat{\mu}_\pm, \hat{K}) - \langle I_\text{int}\rangle(\hat{\mu}_\pm, \hat{K}) \; 
\end{equation}
with $\langle I_\text{int}\rangle (\hat{\mu}_\pm, \hat{K}) = \sum_{i \in \pm} \pi^\mathrm{st}_i I_\text{int}(\hat{\mu}_i, \hat{K})$.
Therefore, the second term in the r.h.s. of the bounds in Eq.~\eqref{eq:MI_bounds} represents a bound on the information interference term emerging from the competing effects of internal interactions and the switching of the mobility matrix.

Eq.~\eqref{eq:xi_hydro} tells us that $\Xi_\mathrm{int, env}$ quantifies how the switching hydrodynamic interactions increase or reduce the total MI with respect to the case in which the environment is static. Since neither the MI nor the dynamic information interference of Gaussian mixtures admits a closed form for all values of $d_\pm$, we rely on Monte Carlo integration with importance sampling to estimate them. In Figure \ref{fig:hydro}e we show that the sign of $\Xi_\mathrm{int, env}$ strongly depends on both the difference between the distances $d_+$ and $d_-$ and the interaction asymmetry $\beta$, with large negative values of $\beta$ favoring constructive interference. Interestingly, $\Xi_\mathrm{int, env}$ tends to vanish when $\beta$ becomes positive. This is due to the fact that, as $\beta$ approaches the edge of instability, $I_{\rm tot}$ becomes dominated by the divergence of the internal MI (see Eq.~\eqref{eq:MI_internal_hydro} and Figure \ref{fig:hydro}b), and thus the interference term becomes negligible.

We note that, in addition to the case of switching distance, several other cases of physically meaningful switching environments are possible. An interesting possibility is the switching of the fluid viscosity $\eta$, which acts as a prefactor in the mobility matrix, and thus appears both in the drift and the noise matrix. As pointed out in Section \ref{sec:slow_jumps} and Appendix \ref{app:fast_switching_prefactors}, this leads to a total MI that coincides with the internal one, since the Lyapunov equation (Eq.~\eqref{eq:Lyapunov_mixture}) no longer depends on the environmental state.

\section{Stochastic systems with\\internal noise}
\noindent Many other stochastic systems, in addition to those discussed above, can be described by Langevin or FP equations with entangled drift and diffusion terms. In fact, this is the usual case when one starts from master equations describing the Markovian dynamics of microscopic processes and derives the corresponding FP equation via a Gaussian noise approximation based on Kramers--Moyal or van Kampen system-size expansions \cite{gardiner}. In this case, the resulting FP equation features drift and diffusion terms that depend on non-trivial combinations of the microscopic parameters. Crucially, these parameters typically appear in both terms, so that random environmental variations -- manifested as temporal fluctuations of the microscopic parameters -- induce simultaneous changes in both the drift and diffusion contributions.

Similar to what was previously observed, the structure of the MI in these systems depends on the specific manner in which the environment enters the dynamics. For fast switching between discrete environmental states, the system dynamics are always governed by the drift and diffusion matrices averaged over the environmental states. In the opposite limit of slow switching, the phenomenology is generally richer and usually leads to information interference terms in the MI.

To provide a concrete and explicit example of how our framework can be implemented, we consider a stochastic chemical reaction network. Starting from a set of microscopic chemical reactions and their associated reaction rates, we write the corresponding master equation and perform linear noise approximation (LNA) to describe deviations from the mean-field solution in the limit of large volume. The resulting drift and diffusion matrices depend on non-trivial combinations of the reaction rates. We then investigate the effect of environmental switching on the MI in this specific setting.

\begin{figure}
    \centering
    \includegraphics[width=\columnwidth]{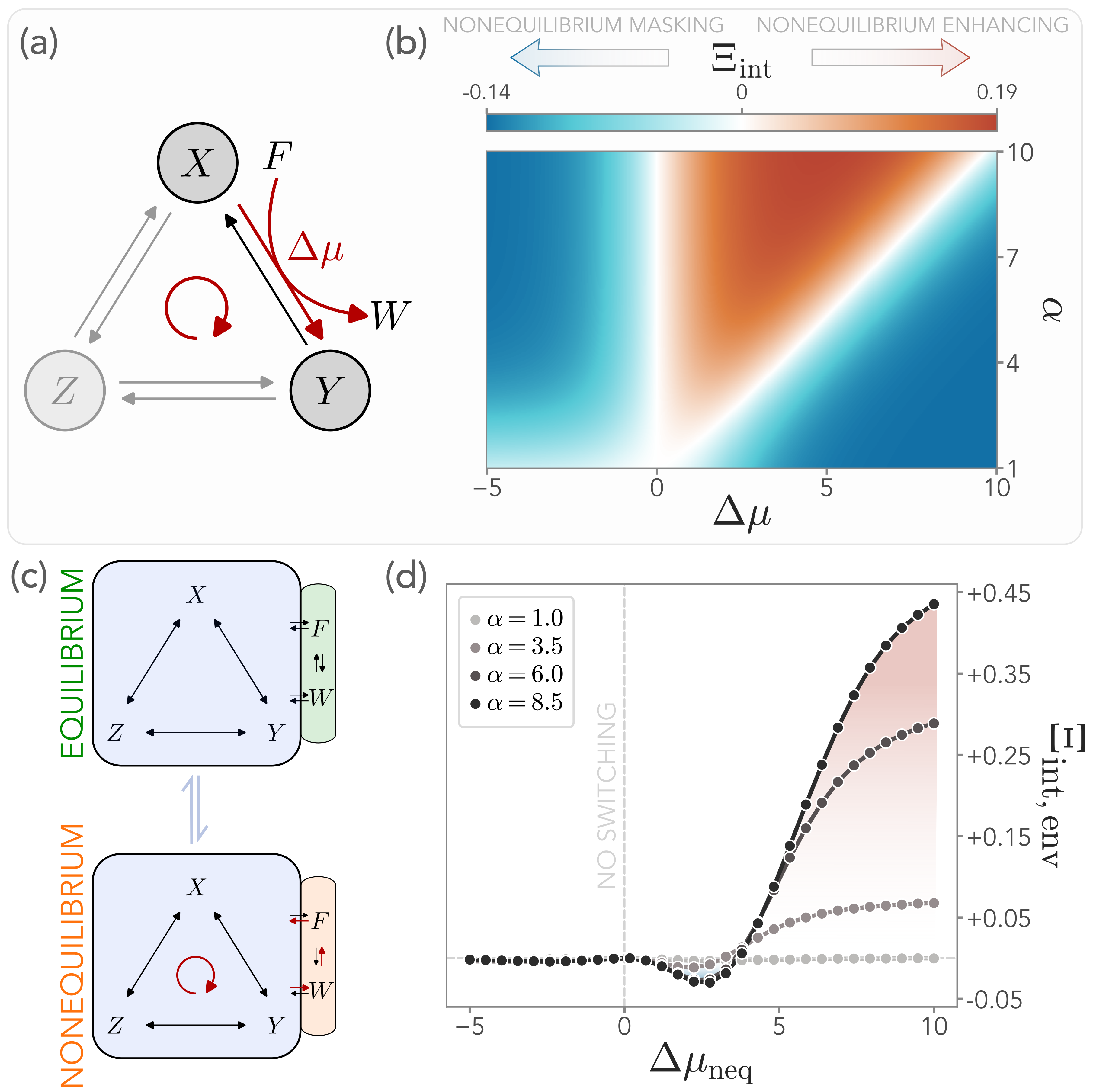}
    \caption{(a) A chemical system with three species $X$, $Y$, and $Z$ that can be driven out of equilibrium by the conversion of chemical fuel ($F$) into waste ($W$), which induces a chemical potential $\Delta \mu$ in the transition from $X$ to $Y$. (b) The noise anisotropy induces a static information interference in the MI between $X$ and $Y$, whose sign is determined by both $\Delta \mu$ and the imbalance in the energy barriers $\alpha$. (c) We consider a changing environment where the system switches between equilibrium and a non-zero chemical potential $\Delta\mu_\mathrm{neq}$. (d) The dynamic information interference term $\Xi_\mathrm{int,env}$ between the environmental MI and the internal one can be both positive or negative if $\Delta\mu_\mathrm{neq} > 0$, showing how the switches to non-equilibrium conditions can both enhance and diminish information. These effects are mostly prominent when $\alpha$ is large. On the other hand, if $\Delta \mu_\mathrm{neq} <0$, corresponding to $[F] < [W]$, the interference term is close to zero, suggesting that the MI is approximately disentangled into the two contributions. Unless otherwise specified, in this Figure $\Delta \epsilon = 1$ and $[W] = 1$.}
    \label{fig:chem}
\end{figure}

\subsection{Fuel-driven chemical cycle}

\par We consider a well-mixed chemical system described by the following reaction cycle (see Fig.~\ref{fig:chem}a):
\begin{gather}
\label{eq:chemical_reactions}
    F + X \xrightleftharpoons[{k_\mathrm{YX}}]{{k_\mathrm{XY}}} Y + W \nonumber \\
    Y \xrightleftharpoons[{k_\mathrm{ZY}}]{{k_\mathrm{YZ}}} Z \qquad Z \xrightleftharpoons[{k_\mathrm{XZ}}]{{k_\mathrm{ZX}}} X \nonumber \;,
\end{gather}
where each $k_{ij}$ denotes a microscopic transition rate and defines the $(ij)$ element of a rate matrix $\hat{\mathcal{K}}$. Here, $F$ and $W$ are a chemical fuel and waste, respectively, and their conversion is associated with the transition from $X$ to $Y$. We consider the concentrations of these external species, $[F]$ and $[W]$, to be chemostatted. Thanks to mass conservation, the total concentration $[X] + [Y] + [Z] = \rho$ is constant. Consequently, the system is fully described by two degrees of freedom, that we arbitrarily choose as being $[X]$ and $[Y]$. These constitute the two coupled stochastic variables under study. Depending on the ratio of $[F]$ to $[W]$, the system can explore both equilibrium and non-equilibrium conditions. For convenience, we define $\Delta \mu = \log([\rm F]/[\rm W])$.

To enforce thermodynamic consistency, we model the transition rates as follows \cite{seifert2012stochastic, rao2016nonequilibrium}:
\begin{gather}
    k_{ij} = k_0 e^{-\beta (B_{ij} - E_i)} \qquad i,j = \rm X,Y,Z \;,
\end{gather}
where $B_{ij} = B_{ji}$ are the energergy barriers that govern the transition from $i$ to $j$, and $E_i$ is the energy of the state $i$. Further employing mass-action kinetics, we obtain the local detailed balance condition for effective rates \cite{rao2016nonequilibrium,peliti2021stochastic,liang2025thermodynamic}:
\begin{equation}
    \label{eq:LDB}
    \frac{\kappa_{ij}}{\kappa_{ji}} = e^{-\beta (E_j - E_i) + \Delta\mu_{ij}} \qquad i,j = \rm X,Y,Z \;,
\end{equation}
where $\kappa_{\rm XY} = k_{\rm XY} [F]$, $\kappa_{\rm YX} = k_{\rm YX} [W]$, while all other effective rates are equal to $k_{ij}$. In Eq.~\eqref{eq:LDB}, $\Delta\mu_{ij} = -\Delta\mu_{ji}$ is the chemical potential difference associated with the transition from $i$ to $j$. In our model, the only non-zero driving terms are $\Delta\mu_{\rm XY} = -\Delta\mu_{\rm YX} = \Delta\mu$. For the sake of simplicity, we take $E_{\rm X} = E_{\rm Y} = E_{\rm Z} = 0$ and $B_{\rm YZ} = \alpha B_{\rm XY}$, with $B_{\rm XZ} = B_{\rm XY} = \Delta \epsilon$.

At equilibrium, the Kolmogorov condition imposes that performing a cycle clockwise or counterclockwise must have the same propensity \cite{schnakenberg1976network,peliti2021stochastic}. Therefore:
\begin{equation}
    \frac{k_{\rm XY} k_{\rm YZ} k_{\rm ZX}}{k_{\rm XZ} k_{\rm ZY} k_{\rm YX}} \frac{[\rm F]^{\rm eq}}{[\rm W]^{\rm eq}} = 1 \to \Delta \mu^{\rm eq} = 0 \;.
\end{equation}
When $\Delta\mu \neq 0$, the system is driven out of equilibrium and sustains a constant non-zero stationary flux. Here we study the system as a function of the energy injected from the chemical fuel and the imbalance in the energy barriers, $\alpha$.

\par As outlined in Appendix \ref{app:chemical_system}, in the limit of large volume the LNA allows to  obtain an appropriate OU equation for the deviation from the mean-field solution. In this equation, the microscopic rates $k_{ij}$ appear in the drift and diffusion matrices in an intricate manner. The internal MI at stationarity reads 
\begin{align}
I_{\rm int}(\hat{\mathcal{K}},&\Delta\mu) = \\&\frac{1}{2}\log\!\left(
\frac{
\left(1+e^{\Delta\mu}\right)
\left(3e^{\Delta\epsilon}+2e^{\alpha\Delta\epsilon}+e^{\Delta\epsilon+\Delta\mu}\right)
}{
\left(2+e^{\Delta\mu}\right)
\left(e^{\Delta\epsilon}+e^{\alpha\Delta\epsilon}+e^{\Delta\epsilon+\Delta\mu}\right)
}
\right) \,, \nonumber
\end{align}
where we have taken $[W] = 1$ for simplicity.

In this scenario, we have two independent processes that contribute to the overall MI between $X$ and $Y$ -- namely, the internal chemical reaction cycle and the fuel-waste conversion that drives the system out of equilibrium. The MI due only to the chemical reaction cycle is obtained by setting $\Delta\mu = 0$ and reads
\begin{equation}
    I_{\rm chem} = \frac{1}{2} \log\left(\frac{4}{3}\right) \; ,
\end{equation}
while the MI due only to fuel-waste conversion, $I_{\rm fw}$, is trivially zero whenever the chemical cycle is shut down, even in the presence of a non-zero driving $\Delta\mu$.
Therefore, following the approach outlined in Section \ref{sec:anisotropy}, the static information interference due to the internal processes reads
\begin{equation}
    \Xi_{\rm int}(\hat{\mathcal{K}},\Delta\mu) =  I_{\rm int}(\hat{\mathcal{K}},\Delta\mu) - I_{\rm chem}
\end{equation}

In Fig.~\ref{fig:chem}b, we show that the presence of a non-equilibrium driving can either mask or enhance the internal MI. In particular, at a fixed $[W]$, $\Delta\mu<0$ effectively suppresses the reaction from $X$ to $Y$, thereby reducing the coupling between the two species. Conversely, a positive $\Delta\mu$ may boost the MI depending on its interplay with the imbalance in the energy barriers, $\alpha$.

We now introduce a switching environment in a situation where the chemical reservoirs that determine $[F]$ and $[W]$ can be controlled externally. In particular, we consider the case in which the chemical system is alternatively exposed to equilibrium and nonequilibrium environments, i.e., $\Delta\mu$ switches between $0$ and a non-zero value $\Delta\mu_{\rm neq}$ 
(see Fig.~\ref{fig:chem}c). As in Eq.~\eqref{eq:MI_decomposition_env}, we compute:
\begin{equation}
    \Xi_{\rm int,env}(\hat{\mathcal{K}},\Delta\mu_{\rm neq}) = I_{\rm tot}(\hat{\mathcal{K}},\Delta\mu_{\rm neq}) - \langle I_{\rm int} \rangle (\hat{\mathcal{K}},\Delta\mu_{\rm neq}).
\end{equation}
Once again, a purely environmental MI does not appear in this case ($I_{\rm env} = 0$) since, by eliminating all internal interactions, which includes chemical interactions, any statistical coupling between $X$ and $Y$ vanishes independently of the value of the driving. In Figure \ref{fig:chem}d, we show that the interference term is much stronger for positive values of $\Delta\mu_{\rm neq}$, but it can be both constructive and destructive. Interestingly, when $\Delta\mu_{\rm neq}<0$, the MI exhibits a ``quasi''-disentangled form, at least for the parameters explored in the example.

\section{Discussion}
\noindent In this work, we have studied how the presence of non-diagonal noise, deterministic couplings, and switching environments affects the statistical dependencies between the DOFs of a system. To quantify such dependencies, we have studied the MI between the DOFs at hand. We have shown that, in general, the total MI cannot be decomposed as the sum of the contributions from deterministic interactions, noise anisotropy, and environmental switching alone. Rather, two sources of information interference emerge. When the environment is static, we have identified an anisotropy-driven information interference term that arises from the interplay between deterministic couplings and the non-diagonal part of the noise. In doing so, we have found that correlated noise fundamentally changes the contribution of deterministic interactions to the statistical dependencies between the DOFs. On the other hand, when the environment changes over time and modifies the parameters of the system, we have introduced a dynamic information interference term that is suitable for studying scenarios in which the environment is much slower than the internal dynamics.

While temperature switches allow for an exact disentangling between the MI induced by the shared environment and the deterministic couplings, the general interplay between the different sources of dependencies is more intricate. In the case of hydrodynamic interactions between two Brownian particles subject to a non-conservative force field, we have found that the presence of a shared fluid can either mask or enhance information. The sign and magnitude of this static information interference between the force and the fluid depend on the degree of non-reciprocity of the force. When the environment induces changes in the equilibrium interparticle distance, the dynamic interference signals that the resulting change in the fluid-mediated interactions can increase the total information. We have also applied these ideas to a fuel-driven chemical system, which can be studied with our framework via an LNA. Interestingly, we have found that the static information interference between the chemical reaction cycle and the fuel-waste conversion process is controlled by the non-equilibrium driving. Furthermore, dynamically switching between the absence of the fuel and a suitable chemical potential -- i.e., between equilibrium and non-equilibrium conditions -- induces a constructive dynamic information interference, which can boost the overall dependencies between the chemical species.

While the framework introduced here provides a systematic decomposition of statistical dependencies in stochastic systems with non-diagonal noise and switching environments, several limitations and open questions deserve attention. First, the analytical results presented here rely on internal linear dynamics or linearization of the dynamics via LNA. Characterizing the information interference framework in genuinely non-linear systems with non-diagonal noise will be important in further work, possibly exploiting analytical tractability in timescale separation regimes \cite{nicoletti2025fast}. Second, as discussed in Appendix \ref{app:MI_reference_frame}, the MI is not invariant under changes of coordinates. Hence, the interference terms defined here depend on the physical parametrization of the system, as it reflects the specific DOFs at hand, and generally requires fixing a natural physical frame.

In future work, it will be interesting to study how information and information interference behave in more complex scenarios. The role of information-theoretic measures in biological systems has been extensively studied, from biochemical to neural networks \cite{tkavcik2011information, bialek2012biophysics, nicoletti2025optimal, tkavcik2016information, mehta2012energetic}. Our framework could provide a natural decomposition of the statistical dependencies arising in these contexts, disentangling the contributions of internal interactions, noise anisotropy, and environmental variability. In particular, it would be interesting to characterize such decomposition in continuously varying environments with anisotropic and spatially heterogeneous diffusion in random media \cite{dorsaz2010diffusion, galanti2014diffusion, chechkin2017brownian}. It would also be interesting to characterize more in general the relation between information interference and non-equilibrium driving in chemical systems. Making this connection quantitative would provide a thermodynamic grounding for the information-theoretic decomposition proposed here. Finally, extending the pairwise framework to systems with more than two degrees of freedom, using multivariate information measures \cite{williams2010nonnegative, rosas2019quantifying, bertschinger2014quantifying}, would allow to study how anisotropy and environmental changes shape dependencies across complex systems.

\appendix
\section{Multivariate Ornstein-Uhlenbeck processes and mutual information}
\label{app:MI_OU}
\noindent The stationary distribution of the process described by Eq. \eqref{eq:Langevin_Ornstein-Uhlenbeck} defined on an unbounded domain is given by the multivariate Gaussian distribution \cite{risken1989fokker}
\begin{equation}
\label{eq:app:p_stat_joint}
    p^\mathrm{st}(\mathbf{x}) = \frac{1}{\sqrt{(2\pi)^N \det \hat{\Sigma}}}\exp\left(-\frac{1}{2}\mathbf{x}^T\hat{\Sigma}^{-1}\mathbf{x}\right),
\end{equation}
with the correlation matrix $\hat{\Sigma}$ that solves the Lyapunov equation (Eq.~\eqref{eq:Lyapunov}). We recall that, given a positive definite matrix $\hat{B} = \hat{b}\hat{b}^T$, there exists a unique positive definite solution $\hat{\Sigma}$ to the Lyapunov equation if and only if $-\hat{A}$ is Hurwitz. 

Since the stationary distribution is a Gaussian, the MI between any pair of degrees of freedom $x_\mu$ and $x_\nu$ (Eq.~\eqref{eq:MI}) can be written as
\begin{equation}
\label{eq:app:MI_Ornstein_Uhlenbeck}
    I_{\mu\nu} = \frac{1}{2}\log\left(\frac{\Sigma_{\mu\mu} \Sigma_{\nu\nu}} {\det\hat{\Sigma}_{\mu\nu}}\right)
\end{equation}
where $\Sigma_{\mu\mu}$ and $\Sigma_{\nu\nu}$ are the $\mu$-th and $\nu$-th diagonal entries of $\hat{\Sigma}$, and $\hat{\Sigma}_{\mu\nu} \in \mathbb{R}^{2\times2}$ is the marginalized covariance matrix of the DOFs at hand.

\section{Environmental switching affecting common scalar prefactor}
\label{app:fast_switching_prefactors}
\noindent In this section, we consider the case where the environmental switching affects a scalar prefactor common to the drift and noise matrices, i.e., $\hat{A}_i = k_i \hat{A}$ and $\hat{B}_i = k_i \hat{B}$. This is for instance the case for switching viscosity in the system of Section \ref{sec:hydro}.

In a fast environment, the Lyapunov equation corresponding to Eq. \eqref{eq:FP_fast_switching} reads
\begin{equation}
    \langle k\rangle\, \hat{A}\,\hat{\Sigma} + \langle k\rangle\, \hat{\Sigma}\,\hat{A}^T = 2\langle k\rangle\,\hat{B} \, ,
\end{equation}
where $\langle k \rangle = \sum_i \pi_i^\mathrm{st}k_i$, whereas in a slow environment Eq.~\eqref{eq:Lyapunov_mixture} becomes
\begin{equation}
    k_i\, \hat{A}\,\hat{\Sigma} + k_i\, \hat{\Sigma}\,\hat{A}^T = 2k_i\,\hat{B} \; .
\end{equation}
In both cases, $\langle k\rangle$ and $k_i$ can be readily simplified, leading to the same solution of an environment-independent covariance matrix. Consequently, the total MI can only depend on the constant matrices $\hat{A}$ and $\hat{B}$, and no environmental information is present. This is not unexpected, as the simplifications in the Lyapunov equation imply that the joint distribution is identical in all environmental states.

\section{Slowly switching temperature}
\label{app:slow_switching_temperature}
\noindent In this section, we consider the case of an environment that switches between $S$ different temperatures $T_i$ with a much longer characteristic timescale than that of the internal dynamics. This can be brought back to Eq. \eqref{eq:Langevin_switch} by defining $\hat{B}_i = T_i\,\hat{B}$. We obtain the set of $S$ Lyapunov equations 
\begin{equation}
\label{eq:Lyapunov_temperature_slow}
    \hat{A}\,\hat{\Sigma}_i + \hat{\Sigma}_i\,\hat{A}^T = 2 \,T_i\,\hat{B},
\end{equation}
which implies that the covariance matrices will be of the form $\hat{\Sigma}_i = T_i\,\hat{\Sigma}$, where $\hat{\Sigma}$ solves a Lyapunov equation that does not depend on the environmental state. Using this fact, the expressions of the Chernoff-$\alpha$ ($C_\alpha$) and Kullback--Leibler ($D_\mathrm{KL}$) divergences for Gaussian distributions \cite{kolchinsky2017estimating} reduce to functions that depend only on the temperatures themselves:
\begin{equation}
\begin{split}
    &C_\alpha (P_{\mu;i} || P_{\mu;j}) = \frac{1}{2}\log\left[\frac{(1-\alpha)T_i + \alpha T_j}{T_i^{1-\alpha}T_j^\alpha}\right]\\
   &C_\alpha (P_{1,\dots,N;i} || P_{1,\dots,N;j}) = S \, C_\alpha (P_{\mu;i} || P_{\mu;j})\\
    &D_\mathrm{KL}(P_{\mu;i} || P_{\mu;j}) = \frac{1}{2}\left[\log\left(\frac{T_j}{T_i}\right) + \text{Tr}\left(\frac{T_i}{T_j}\right) - 1\right]\\ &D_\mathrm{KL}(P_{1,\dots,N;i} || P_{1,\dots,N;j}) = S \, D_\mathrm{KL}(P_{\mu;i} || P_{\mu;j})
\end{split}
\end{equation}
where $\alpha \in [0,1]$ is arbitrary. To minimize the Chernoff-$\alpha$ divergence, we choose $\alpha$ such that $\partial C_\alpha /\partial \alpha = 0$, obtaining
\begin{equation}
    \alpha = \frac{T_i}{T_i - T_j} + \log\left(\frac{T_i}{T_j}\right).
\end{equation}
With this choice, the Chernoff-$\alpha$ divergence becomes a function only of the set of ratios $\gamma_{ij} \equiv T_i/T_j$:
\begin{equation}
     C_\alpha (P_{\mu;i} || P_{\mu;j}) = \frac{1}{2} \left[\log\left[\frac{(\gamma_{ij}-1) \gamma_{ij}^{\frac{1}{\gamma_{ij}-1}}}{\log\gamma_{ij}}\right]-1\right].
\end{equation}
Analogously, the Kullback-Leibler divergence is also a function of the $\{\gamma_{ij}\}$ only:
\begin{equation}
    D_\mathrm{KL}(P_{\mu;i} || P_{\mu;j}) = \frac{1}{2}\left(\gamma_{ij} - \log\gamma_{ij} - 1\right) \; .
\end{equation}
This result highlights that the second term of the r.h.s. of Eq.~\eqref{eq:MI_bounds} is a function only of the ratios $\gamma_{ij}$ and the switching rates. 

On the other hand, the fact that $\hat{\Sigma}_i = T_i\hat{\Sigma}$ implies that the MI in Eq.~\eqref{eq:app:MI_Ornstein_Uhlenbeck} does not depend on the temperature. Hence, the internal MI in this case is only a function of $\hat{A}$ and $\hat{B}$ in all environmental states, so that
\begin{equation}
\label{eq:app:MI_slow_temperature_internal}
    \langle I_\text{int}(\hat{A},\hat{B})\rangle = I_\text{int}(\hat{A},\hat{B})
\end{equation}
is independent of the environment. Thus, in a system with slowly switching temperatures, the bounds on the MI are effectively disentangled:
\begin{equation}
\label{eq:app:MI_bounds_temperature}
    I_\text{tot}^\text{slow,up/low} = I_\text{int}(\hat{A}, \hat{B}) + I_\text{env}^\text{up/low}\left(\{\gamma_{ij}\}, \{\pi_i\}\right)
\end{equation}
where $\pi_i$ is the probability of being in the $i$-th environmental state, so that the MI depends in general on ratios of elements of the rate matrix $\hat{W}$ that governs environmental switches. Importantly, the environmental term of the upper and lower bounds converge to a common value when all the $S$ temperatures are similar ($\gamma_{ij} \to 1 ~\forall\, i,j$) or diverge ($\gamma_{ij} \to 0$ or $\gamma_{ij}\to\infty$, $\forall\, i,j$). In the former case, all Cernoff-$\alpha$ and Kullback--Leibler divergencies tend to zero; in the latter case, they all tend to infinity. Consequently,
\begin{equation}
    \begin{split}
        \lim_{\gamma_{ij} \to 1} I_\text{env}^\text{up/low}(\{\gamma_{ij}\}) &= 0\\
        \lim_{\gamma_{ij} \to 0,\infty} I_\text{env}^\text{up/low}(\{\gamma_{ij}\}) 
        &= (S -1) H_\text{jumps}
    \end{split}
\end{equation}
where
\begin{equation}
    H_\text{jumps} \equiv - \sum_i \pi_i \log \pi_i
\end{equation}
is the Shannon entropy of the environment. In these limits, the total MI is exactly disentangled into an internal and an environmental term. When $S = 2$, Eq. \eqref{eq:MI_slow_temperature_limits} is recovered.

We briefly note here that the internal MI vanishes if and only if $\hat{A}$ and $\hat{B}$ are both diagonal, indicating null internal couplings and noise correlations. Indeed, in this case the dynamics is reversible, and the covariance matrix is \cite{Godreche_2019}
\begin{equation}
    \hat{\Sigma}_i = T_i \hat{A}^{-1} \hat{B},
\end{equation}
which is diagonal as well. This implies that $\det \hat{\Sigma_i} = T_i\,\prod_\mu {\Sigma}_{\mu\mu}$ and the logarithm in Eq. \eqref{eq:app:MI_Ornstein_Uhlenbeck} vanishes. Therefore, when this occurs, we can identify the total MI with the environmental term, i.e., the MI in the absence of deterministic couplings.

Finally, we briefly comment on the opposite case where the environment changes the drift matrix as $\hat{A}_i = k_i \hat{A}$ but leaves the noise matrix unchanged. In a slow environment, the Lyapunov equation (Eq.~\eqref{eq:Lyapunov_mixture}) becomes
\begin{equation}
    \hat{A}\,\hat{\Sigma} + \hat{\Sigma}\,\hat{A}^T = 2\frac{1}{k_i}\hat{B} \; .
\end{equation}
Thus, we immediately recover the same phenomenology of the slowly-switching temperatures by identifying $k_i$ with an inverse temperature. We note that a similar result can be obtained for fast environments, where $\langle T \rangle = \sum_i \pi_i^\mathrm{st} /k_i$.

\section{Derivations for the chemical system}
\label{app:chemical_system}

\par Let us start from Eq. \eqref{eq:chemical_reactions}, denote the vector of concentrations by $\mathbf{x} = (x,y) = ([X],[Y])$ and the system volume by $V$. Under the assumption of a Markovian dynamics, the system is fully described by the master equation
\begin{equation}
\label{eq:master_equation}
    \dot{P}(\mathbf{x},t) = V \sum_\mathbf{x'}\Big[T(\mathbf{x}|\mathbf{x'})\,P(\mathbf{x'},t) - T(\mathbf{x'}|\mathbf{x})\,P(\mathbf{x},t)\Big]
\end{equation}
with $T(\mathbf{x'}|\mathbf{x})$ being the density-dependent macroscopic transition rates from state $\mathbf{x}$ to state $\mathbf{x'}$ and given by
\begin{equation}
    \begin{split}
        T(x-1/V, y+1/V|x,y) &= k_\mathrm{XY}\, x\\
        T(x+1/V, y-1/V|x,y) &= k_\mathrm{YX}\, y\\
        T(x, y-1/V|x,y) &= k_\mathrm{YZ}\, y\\
        T(x, y+1/V|x,y) &= k_\mathrm{ZY}\, (\rho - x - y)\\
        T(x+1/V, y|x,y) &= k_\mathrm{ZX}\, (\rho - x - y)\\
        T(x-1/V, y|x,y) &= k_\mathrm{XZ}\, x\\
    \end{split}
\end{equation}

\par Let $A_i(\mathbf{x})$ and $B_{ij}(\mathbf{x})$ be the first- and second-order Kramers--Moyal coefficients associated with the master Eq. \eqref{eq:master_equation}. Specifically, they read
\begin{equation}
\begin{split}
    A_i(\mathbf{x}) = V &\sum_{\mathbf{x'}} (x_i' - x_i) \,T(\mathbf{x'}|\mathbf{x}),\\
    B_{ij}(\mathbf{x}) = \frac{V^2}{2} &\sum_{\mathbf{x'}} (x_i'- x_i)(x_j'- x_j) \,T(\mathbf{x'}|\mathbf{x}).
\end{split}
\end{equation}
When the volume is large, we can perform the LNA \cite{gardiner}. We make the ansatz
\begin{equation}
\label{eq:LNA_ansatz}
    \mathbf{x} = \boldsymbol{\phi}(t) + \frac{1}{\sqrt{V}}\boldsymbol{\xi},
\end{equation}
where $\boldsymbol{\phi}$ is the deterministic solution and $\boldsymbol{\xi}$ is a stochastic variable that describes the fluctuations around $\boldsymbol{\phi}$. We then substitute Eq. \eqref{eq:LNA_ansatz} into the master Eq. \eqref{eq:master_equation} and expand in powers of $V$. At first order we obtain the deterministic equation
\begin{equation}
\label{eq:LNA_deterministic}
    \dot{\boldsymbol{\phi}}(t) = \mathbf{A}(\boldsymbol{\phi}),
\end{equation}
while at second order we obtain an FP equation for the probability distribution of $\boldsymbol{\xi}$:
\begin{equation}
\label{eq:LNA_FP}
    \dot{p}(\boldsymbol{\xi},t) = \nabla\cdot\left[-\hat{A'}(\boldsymbol{\phi})\,\boldsymbol{\xi}\,p(\boldsymbol{\xi},t) + \hat{B}(\boldsymbol{\phi})\,\nabla p(\boldsymbol{\xi},t)\right],
\end{equation}
where $\hat{A'}(\boldsymbol{\phi})$ denotes the Jacobian of $\mathbf{A}(\boldsymbol{\phi})$.

\par Let $\boldsymbol{\phi^*}$ be the stationary solution of Eq. \eqref{eq:LNA_deterministic}. If $\hat{A'}(\boldsymbol{\phi^*})$ is Hurwitz, the equilibrium is stable. Since Eq. \eqref{eq:LNA_FP} describes an OU process, the stationary distribution of $\boldsymbol{\xi}$ is given by a Gaussian whose covariance matrix solves the Lyapunov equation (see Appendix \ref{app:MI_OU})
\begin{equation}
    \hat{A'}(\boldsymbol{\phi^*})\hat{\Sigma} + \hat{\Sigma}[\hat{A'}(\boldsymbol{\phi^*})]^T = -2\hat{B}(\boldsymbol{\phi^*}).
\end{equation}
In the given chemical system, the two matrices read

\begin{equation}
\resizebox{0.45\textwidth}{!}{$
        \hat{A'}(\boldsymbol{\phi^*}) = \begin{pmatrix}
            - k_\mathrm{XY} - k_\mathrm{XZ} - k_\mathrm{ZX}
&
k_\mathrm{YX} - k_\mathrm{ZX}
\\[0.8em]
k_\mathrm{XY} - k_\mathrm{ZY}
&
- k_\mathrm{YX} - k_\mathrm{YZ} - k_\mathrm{ZY}
\end{pmatrix}
$}
\end{equation}
and
\begin{widetext}
\begin{equation}
            \resizebox{\textwidth}{!}{$
        \hat{B}(\boldsymbol{\phi^*}) = \dfrac{\rho}{2D}\begin{pmatrix}
2 (k_\mathrm{XY}+k_\mathrm{XZ})
\left[k_\mathrm{ZX}(k_\mathrm{YX}+k_\mathrm{YZ}) + k_\mathrm{YX}k_\mathrm{ZY}\right]
&
-\left[
k_\mathrm{YX}k_\mathrm{ZY}(2k_\mathrm{XY}+k_\mathrm{XZ})
+ k_\mathrm{XY}k_\mathrm{ZX}(2k_\mathrm{YX}+k_\mathrm{YZ})
\right]
\\[1.2em]
-\left[
k_\mathrm{YX}k_\mathrm{ZY}(2k_\mathrm{XY}+k_\mathrm{XZ})
+ k_\mathrm{XY}k_\mathrm{ZX}(2k_\mathrm{YX}+k_\mathrm{YZ})
\right]
&
2 (k_\mathrm{YX}+k_\mathrm{YZ})
\left[k_\mathrm{ZY}(k_\mathrm{XY}+k_\mathrm{XZ}) + k_\mathrm{XY}k_\mathrm{ZX}\right]
\end{pmatrix}
    $},
\end{equation}
\end{widetext}
with denominator
\begin{equation}
\begin{split}
    D &= (k_\mathrm{XY}+k_\mathrm{XZ})(k_\mathrm{YZ}+k_\mathrm{ZY})
+ \\
&+ k_\mathrm{ZX}(k_\mathrm{XY}+k_\mathrm{YX}+k_\mathrm{YZ}) +\\
&+ k_\mathrm{YX}(k_\mathrm{XZ}+k_\mathrm{ZY}).
\end{split}
\end{equation}
This is therefore another example in which the two stochastic variables $\xi_x$ and $\xi_y$ are coupled through both the deterministic term and a non-diagonal noise. In addition, the microscopic parameters enter the drift and noise matrix in an intricate manner, so that any environmental changes would produce complex simultaneous variations of the two matrices.

\section{Frame-dependence of mutual information}
\label{app:MI_reference_frame}
\noindent In this Appendix, we briefly recall the consequences for the MI of the non-invariance under reparametrization of the Kullback-Leibler divergence. Let us consider the case in which $\hat{b}$ is real and symmetric. The spectral theorem ensures that there exists an orthogonal matrix $\hat{P}$, whose columns are given by the eigenvectors of $\hat{b}$, such that
\begin{equation}
    \hat{b} = \hat{P}\hat{\Lambda}_b\hat{P}^T
\end{equation}
with $\hat{\Lambda}_b$ being a diagonal matrix. Using that $\hat{P}\hat{P}^T = \hat{P}^T\hat{P} = \mathbb{1}$ and defining $\mathbf{y} = \hat{P}^T \mathbf{x}$ and $\bm{\eta} = \hat{P}^T \bm{\xi}$, we can rewrite the Langevin Eq. \eqref{eq:Langevin_Ornstein-Uhlenbeck} as
\begin{equation}
    \dot{\mathbf{y}} = -\hat{P}^T\hat{A}\hat{P}\mathbf{y} + \sqrt{2}\hat{\Lambda}_b\bm{\eta},
\end{equation}
where $\bm{\eta}$ is again delta correlated. From Eq.~\eqref{eq:MI_decomposition_anisotropy} one can deduce that, in the new reference frame, the MI depends only on the deterministic dynamics ($I_\text{int} = I_\text{coup}$). Nevertheless, this in general does not equate  the original MI from noise couplings. For example, take $N = 2$, $\hat{A} = \mathbb{1}$ and $\hat{b}$ as in Eq. \eqref{eq:A_b}. Before transformation, the MI comes purely from noise and is given by Eq. \eqref{eq:MI_noise}. However, after transformation, and since $\hat{P}^T\hat{A}\hat{P} = \mathbb{1}$, the MI is null. On the other hand, if $\hat{b}$ is non-symmetric, the matrix $\hat{P}$ is not orthogonal, and the transformed noise $\bm{\eta}$ will not be delta correlated. In fact, in this case $\bm{\eta} = \hat{P}^{-1}\bm{\xi}$ and
\begin{equation}
    \langle \bm{\eta} \bm{\eta}^T \rangle = \hat{P}^{-1} \langle \bm{\xi} \bm{\xi}^T \rangle(\hat{P}^{-1})^T =  \hat{P}^{-1}(\hat{P}^{-1})^T,
\end{equation}
which in general is not equal to the identity matrix. In this case, it is not possible to transfer noise couplings to the drift term by a change of variable.

This implies that, in general, the MI is frame-dependent. In particular, there will always be a coordinate system in which the MI is null. Indeed, since $\hat{\Sigma}$ is positive definite, its inverse $\hat{\Sigma}^{-1}$ is also positive definite, and one can always define a change of variables that diagonalizes it. In this case, the stationary joint probability of Eq. \eqref{eq:app:p_stat_joint} can be factorized exactly, leading to a vanishing logarithmic term in Eq. \eqref{eq:MI}.  In general, however, such transformation does not render the noise matrix diagonal, and the MI is null for a specific redistribution of couplings between the noise and the deterministic terms.

\bibliography{references}

\end{document}